\documentclass[12pt,preprint]{aastex}

\shortauthors{Cao \& Yu}

\begin{document}

\title{{Frequency Variation of the Kilohertz Quasi-periodic Oscillations and the Flux of the Band-limited Noise in Scorpius X-1}}

\author{Xiaofeng Cao\altaffilmark{1,2} and Wenfei Yu\altaffilmark{1}}
\altaffiltext{1}{Shanghai Astronomical Observatory, 80 Nandan Road, Shanghai 200030, China; wenfei@shao.ac.cn }
\altaffiltext{2}{The Institute of Astrophysics, Huazhong Normal University, Wuhan 430079, China}

\begin{abstract}
We study the kilohertz quasi-periodic oscillations (kHz QPOs) and the band-limited noise (BLN) in the 0.5--16 Hz range observed simultaneously on the horizontal branch (HB) and on the upper normal branch (NB) of the brightest neutron star Low-mass X-ray Binary (LMXB) Scorpius X--1 with the observations performed with the {\it Rossi X-Ray Timing Explorer (RXTE)}.  We find that the twin kHz QPO frequencies are positively correlated with the flux variations taking place on the BLN time scales on the HB, in contrast to the anti-correlation held on the time scale of the normal branch oscillation (NBO) on the NB reported previously, suggesting that although they occur in sequence along the color-color tracks, the BLN and the NBO are of different origins. We also show the evidence that the frequency separation between the twin kHz QPOs decreases with the flux by $2\sim~3$ Hz on the BLN time scales, which is consistent with the trend on the longer time scale that the Z source traces the HB. This further suggests that the flux variation associated with the BLN originates from the mass accretion rate variation in the disk accretion flow. We discuss the implications of these results for our understanding of the BLN. 
\end{abstract}


\section{INTRODUCTION}
The brightest neutron star Low-mass X-ray Binaries (LMXBs) were divided into two main types, namely,  the ``Z" sources and the ``atoll" sources, based on their spectral and timing behavior studied twenty years ago (Hasinger \& van der Klis 1989). The brighter type is the ``Z" sources, because they trace out roughly a Z-shape in X-ray color-color diagram (CD), which consists of three branches, called the horizontal branch (HB), the normal branch (NB), and the flaring branch (FB), respectively. The picture is yet not well understood. Recent observations of the transient Z source XTE J1701-462 have brought new clues to the formation of the tracks (Homan et al. 2007), while 
observations in the past decade revealed some weak neutron star LMXBs accreting at lower mass accretion rates, some of which show similarities to the atoll sources (see the review van der Klis 2006). 

The power spectra of neutron star LMXBs usually contain several variability components in a wide range of frequency up to kilohertz. Aperiodic variability includes Quasi-periodic Oscillation (QPO, narrow feature) or noise (broad structure). Band-limited noise (BLN) component, formerly called high-frequency noise (HFN) in atoll sources and low-frequency noise (LFN) in Z sources, has a flat shape up to a break frequency. For the Z sources, the  BLN is observed only on the HB and the very upper part of the NB. In addition to the BLN, various QPOs are observed in different branches of the Z sources. The characteristic frequencies of the variability components often vary monotonically along the tracks in the CD.

The brightest persistent neutron star LMXB \mbox{Sco X$-$1} is a Z source. In \mbox{Sco X$-$1} there are three distinct types of QPOs observed, i.e., the near 6 Hz normal-branch oscillation (NBO), the near 45 Hz horizontal-branch oscillation (HBO) with a harmonic near 90 Hz, and the kilohertz quasi-periodic oscillations (kHz QPOs). The kHz QPOs and the harmonic of the HBO, were first seen in the persistent flux of \mbox{Sco X$-$1} from the very early RXTE observations (van der Klis et al. 1996a; 1997). 

The characteristic frequencies of the X-ray variability in X-ray binaries are probably related to some characteristic time scales in the system, some of which may be used to constrain the mass and the spin of the compact object. Many authors have tried to explain these  frequencies and their properties. Models for the kHz QPOs, which are based on orbital and epicyclical motions, include relativistic precession model (Stella \& Vietri 1999) and sonic-point and spin-resonance model (Miller et al. 1998; Lamb \& Miller 2004). Some other models for the kHz QPOs are based on disk oscillations or modes in the accretion flow, e.g., nonlinear resonance (Abramowicz \& Kluzniak 2001; Abramowicsz 2005), Alfve\'{n} wave oscillation (Zhang 2004), and kink mode (Li \& Zhang 2005). Models for the HBOs include magnetospheric beat-frequency models (Alpar \& Shaham 1985; Lamb et al. 1985), the Lense-Thirring precession model (Stella \& Vietri 1998), and the magnetically warped precessing-disk model (Lai 1999; Shirakawa \& Lai 2002). For the NBO or the FBO, disk oscillation (e.g. Alpar et al. 1992), radiation-force feedback instability (Fortner et al. 1989), and the oscillation of a spherical shell (Titarchuk 2001) have been proposed. Related to \mbox{Sco X$-$1}, a possible explanation for the NBO and the coupling between the twin kHz QPOs and the NBO was discussed by Hor$\acute{a}$k et al. (2004). In addition, a two-oscillator (TO) model has been developed to explain several  characteristic frequencies in neutron star X-ray binaries (Titarchuk \& Osherovich 1999; Titarchuk et al. 1999). 

The frequency correlation between these QPOs and noises has been found to occur in different source types and over a wide range of luminosities (e.g., Psaltis, Belloni, and van der Klis 1999; Belloni, Psaltis, and van der Klis 2002; van der Klis 2006). The correlations between the frequencies of the QPOs and the X-ray count rate on long-term and short-term time scales were observed as well (note: past X-ray timing measurements we discussed below were obtained with two proportional counter instruments, i.e., the Medium Energy (ME) Proportional Counter on board the European Space Agency's X-ray Observatory (EXOSAT) and the Proportional Counter Array (PCA) on board the RXTE. Therefore the correlations between the frequencies of various QPOs and the X-ray count rate as seen with the EXOSAT and the RXTE are generally consistent). In the atoll sources on the time scales of hours to days, the kHz QPO frequency is found correlated with the X-ray count rate, but on time scales longer than days the correlation does not hold and we see the ``parallel track'' phenomenon (e.g., \mbox{4U~0614+091}, Ford et al. 1997; \mbox{4U 1608$-$52}, Yu et al. 1997; M\'{e}ndez et al. 1999). An anti-correlation between the kHz QPO frequency and the count rate associated with the mHz QPO, which might correspond to a thermonuclear burning mode on neutron star, was also found in 4U 1608-52 (Yu \& van der Klis 2002). In the Z sources on the time scales of hours to days, the kHz QPO frequencies and the X-ray count rate appear to positively correlate on the HB and sometimes anti-correlate on the NB as inferred from observations reported in the literature. In \mbox{GX 5$-$1} and \mbox{GX 17+2}, the kHz QPO frequencies increase from the HB to the upper NB (van der Klis et al. 1996b), while the X-ray count rate increases from the HB to the HB/NB vertex and then decreases along the upper NB (Kuulkers et al. 1994; Wijnands et al. 1997; Homan et al. 2002), therefore a positive correlation on the HB and an anti-correlation on the upper NB can be inferred. In \mbox{GX 340+0} and \mbox{Cyg X$-$2}, the kHz QPO frequencies and the count rate increase along the HB to the HB/NB vertex (Jonker et al. 1998; Wijnands et al. 1998), so there is a positive correlation on the HB. In \mbox{Sco X$-$1}, the kHz QPO frequencies increase on the NB (van der Klis et al 1996a) and the X-ray count rate decreases from the HB to the NB (Dieters \& van der Klis 2000), so there is an anti-correlatation on the NB (Yu et al. 2001). Similar to those of the kHz QPOs, the frequency of the HBO is correlated with the X-ray count rate as well. In general, the HBO frequency and the count rate appear to positively correlate on the HB. For example, in \mbox{GX 5$-$1}, the HBO frequency is positively correlated with the source count rate on the HB (Lewin et al.1992; Kuulkers et al. 1994). But the situation is complex on the NB. In \mbox{Cyg X$-$2}, the HBO frequency increases from the HB to the HB/NB vertex and then remains approximately constant on the upper NB, while the count rate increases from the HB to the very upper NB (Wijnands et al. 1998). In \mbox{GX 340+0}, the HBO frequency increases from the HB to the HB/NB vertex and then remains approximately constant on the upper NB, while the count rate increases from the HB to the HB/NB vertex and decreases from the NB to the NB/FB vertex (Kuulkers \& van der Klis 1996; Jonker et al. 1998). In \mbox{GX 17+2}, the HBO frequency increases from the HB to the middle NB and then decreases from the middle NB to the NB/FB vertex, while the count rate increases from the HB to the HB/NB vertex and decreases from the NB to the NB/FB vertex (Wijnands et al. 1996; Homan et al. 2002). Therefore on the upper NB, previous observations suggest that there is no significant correlation between the HBO frequency and the X-ray count rate in \mbox{Cyg X$-$2} and \mbox{GX 340+0}, while there is an anti-correlation in GX 17+2. On the lower NB, a positive correlation was seen in \mbox{GX 17+2}. Similar to the atoll sources, on longer time scales, there is no correlation between the QPO frequencies and the X-ray count rate. 

Previous studies show the BLN on the HB and the NBO on the NB in \mbox{Sco X$-$1} occur in time and frequency sequence (see Figure 2 of van der Klis et al. 1997). One may wonder if they are of similar origins. In this paper, we investigated the dependence of the kHz QPOs on the flux variation due to the BLN in \mbox{Sco X$-$1}. We found that on the time scales of the BLN, there is a positive correlation between the kHz QPO frequencies and the X-ray flux , and the frequency separation of the kHz QPOs deceases with the X-ray flux, which is different from what was seen for the NBO.

\section{Observations and data analysis}
The observations of \mbox{Sco X$-$1} performed with the Proportional Counter Array (PCA) on board the {\it Rossi X-Ray Timing Explorer} ({\it RXTE}; Bradt et al. 1993) between 1996 February 14 and 2006 September 28 were used in our preliminary analysis. The high time resolution data mode is Bin mode, including B$\_$125us$\_$1M$\_$0$\_$87$\_$H, B$\_$250us$\_$1M$\_$0$\_$87$\_$H, B$\_$250us$\_$1M$\_$0$\_$249$\_$H, B$\_$250us$\_$2A$\_$0$\_$49$\_$H, or B$\_$250us$\_$2A$\_$8$\_$39$\_$H, which were used to generate the power spectrum of each observation. Data from all active Proportional Counter Units (PCUs), ranging from 3 to 5, were extracted over the whole energy range available in each data modes and rebinned to 1/{4096} s resolution throughout.

\subsection{Study of BLN on the HB}
First, we made fast Fourier transform with lengths of 32 s to search for the observations with an average power spectrum composed of a BLN and twin kHz QPOs. We wanted to select the observations showing significant twin kHz QPOs and strong BLN on the HB. The flat top of the BLN in the observations on the HB is in the frequency range around 0.5 -- 4.0 Hz, so we calculated the average power in the frequency range 0.5 -- 4.0 Hz for each of these observations and picked out those with an average power greater than 5.0 (Leahy normalization). Then for each of these observations, we used transform lengths of $1/{32}$ s to generate a 32 Hz resolution average power spectrum with a Nyquist frequency of 2048 Hz. We fitted each average power spectrum in the 192--2048 Hz range with a model of the form $P(\nu)=a_1 + a_2\nu^{a_3}$, where $a_1, a_2, a_3$ are constants, to account for the dead-time-modified Poisson noise and two Lorentzian peaks for the twin kHz QPOs at high frequencies. We determined the centroid frequencies ($\nu_{2}$ for the upper and $\nu_{1}$ for the lower), the corresponding integrated powers, and their uncertainties of the kHz QPOs. We then picked out the observations in which the upper kHz QPO is strong with an integrated Leahy normalized power greater than 18. The observation IDs of the selected 5 observations are listed in Table 1 (group A). A typical average power spectrum corresponding to these observations is shown in Figure 1. It shows that the power spectrum primarily includes a BLN with a characteristic frequency around a few Hz and twin kHz QPOs. A model composed of a Lorentzian for the BLN at low frequencies, two Lorentzians for the twin kHz QPOs at high frequencies, and a power-law noise plus a constant to account for the dead-time affected Poisson noise is shown in the plot as well. 

With the five observations, we intended to study the relation between the frequencies of the twin kHz QPOs and the flux variation due to the BLN. We had to average power spectra from different observations in which the frequencies of the kHz QPOs are different. Therefore we need to shift the power spectra in frequency to align the kHz QPO peaks so that we can detect QPO frequency drift when the BLN flux varies. First, we determined the frequency shift for $\nu_{2}$ in the power spectrum of each observation to be aligned at 800 Hz. This frequency shift was used later on. We sampled the power spectra according to the BLN flux. We used transform lengths of $1/{32}$ s to calculate a series of Fourier power spectra for each observation. Similar to the method used by Yu et al. (2001) but without a selection on the significance of the count rate difference, the series of power spectra were assigned into many non-overlapping pairs of successive power spectra. We compared the two average count rates of each pair of successive power spectra and constructed the following two groups of power spectra. The high flux group includes the one with a higher rate in each pair, and the low flux group includes the lower rate one. The average count rate difference ($C_{H}-C_{L}$, where $C_{H}$ and $C_{L}$ denote high and low rate respectively) between the two groups describes the flux variation at the variability frequency  close to 16 Hz. After shifted by the determined frequency shift to align the upper kHz QPO at 800 Hz, the power spectra of the two groups were averaged respectively to generate two average power spectra. We fitted both average power spectra in the 192--1920 Hz range with two Lorentzian peaks and a curve of the form $P(\nu)=a_1 + a_2\nu^{a_3}$ to characterize the power spectra, and determined the central frequencies of the kHz QPOs ($\nu_{1H}$ and $\nu_{2H}$ for high flux group; $\nu_{1L}$ and $\nu_{2L}$ for low flux group). We then compared the centroid frequencies of the kHz QPOs between the two groups, and calculated frequency differences of the upper and the lower kHz QPOs ($\nu_{2H}-\nu_{2L}$ and $\nu_{1H}-\nu_{1L}$), respectively.

The difference between the average count rates of the high flux group and the low flux group is related to the Fourier power around frequency $1/{2\delta t}$. Hence, with segment duration $\delta t=1/{32}$ s, we sampled the flux variation near 16 Hz. The BLN dominates the power spectrum in the frequency range from about 0.5 Hz to 16 Hz. In order to cover this range, we constructed longer segments of duration $N\delta t$, where $N=2,3, ..., 32$, which were used to sample variations on the time scales corresponding to $16/N$ Hz. We determined the corresponding count rates and the corresponding  power spectra for each N. With these power spectra, we constructed high count rate and low count rate groups of power spectra and compared the QPO frequencies in the same way as above for $N=1$. For most $N$, $\nu_1$ and $\nu_2$ both are higher in the high count rate group than in the lower count rate group. The frequency drifts of $\nu_1$ and $\nu_2$ are shown in Figure 2b and 2d as filled circles, respectively. We also show the frequency drifts over count rate difference in units of counts per 1/32 s in Figure 2c and 2e.

To study QPO frequency drifts in response to flux variations at lower frequencies than 0.5 Hz, we made fast Fourier transforms with segment duration $\delta t=0.5$ s, which resulted in power spectra with 2 Hz resolution, higher than the study of segments with a duration of $\delta t=1/{32}$ s. With these power spectra, we can sample fluctuations near 1/M Hz in the range from 0.03 Hz to 1.0 Hz with a frequency resolution 16 times smaller, where $M=1,2,3, ..., 32$. The  results are shown in Figure 2 as diamonds.

For the upper kHz QPO frequency $\nu_2$, the average frequency drift corresponding to the flux variation of the BLN between 0.5 Hz and 16 Hz is $1.8\pm0.4$ Hz; for the lower kHz QPO frequency $\nu_1$, the average frequency drift is $4.4\pm0.6$ Hz. It appears that the frequency drift in $\nu_1$ is larger than that of  $\nu_2$ and the peak separation $\Delta\nu$ of the twin kHz QPOs is $2.6\pm0.7$ Hz smaller for the high count rate group than that for the low count rate group. For the flux variation in the frequency range 0.03--1.0 Hz, the average frequency shifts of $\nu_2$ and $\nu_1$ are $0.9\pm0.3$ Hz and $1.6\pm0.6$ Hz, respectively, which are not as significant as those corresponding to the flux variation in the frequency range 0.5--16 Hz. This is reasonable since most of the flux variation due to the BLN is in the 1--10 Hz range. 

A QPO frequency drift on longer time scales could contribute to low level of QPO frequency drift on the BLN time scales. Therefore,  we compared the correlation between the kHz QPO frequency and the flux variation on the BLN time scales with that on longer time scales to judge whether the QPO frequency drift on the BLN time scales is the result of the QPO frequency drift on longer time scales. We divided the data of the five observations into segments of length 256 s. We calculated the power spectrum of each segment and determined the frequencies of the kHz QPOs and the corresponding count rate. Figure 3a and 3b shows $\nu_1$ and $\nu_2$ against the count rate per PCU on 256 s time scale  for all selected observations. We found that on the 256 s time scale there is no apparent correlation between the frequencies of the twin kHz QPOs and the X-ray count rate. There might be a positive correlation in the observation 20053-01-01-00. We then studied whether the positive correlation on the BLN time scale we got came from 20053-01-01-00 alone. We found the lower and the upper kHz QPO frequencies drifted $2.4\pm0.6$ Hz and $4.7\pm1.0$ Hz for 20053-01-01-00 and $1.5\pm0.5$ Hz and $4.5\pm0.8$ Hz for the other four observations, respectively, suggesting that the positive correlation on the BLN time scales was not from the effect of variations on longer time scales. 

\subsection{ Study of BLN on the NB}
It is known that in \mbox{Sco X$-$1} the BLN shows up on the upper NB as well. We therefore  further studied the relation between the frequencies of the twin kHz QPOs and the flux variation due to the BLN evolving from the HB to the upper NB. The segments of the observations of \mbox{Sco X$-$1} evolving from the HB to the upper NB during which \mbox{Sco X$-$1} show strong  BLN and twin kHz QPOs in the power spectra were studied. The strength of the BLN and the kHz QPOs in each observation were determined in the same way as our previous analysis. We selected the observation segments during which the upper kHz QPO had a Leahy normalized integrated power greater than 4  and the average power of BLN was greater than 2 and  less than 5 in the frequency range 0.5 -- 4.0 Hz ( BLN with an average power greater than 5 corresponds to the HB case studied above). We made fast Fourier transform with intervals of 32 s for each continuous observation segments in these observations. We constructed three groups of segments, namely the B, C and D group,  according to the BLN strength with an average power in the range 4.0--5.0, 3.0--4.0, and 2.0--3.0, respectively.The groups of observation segments are listed in Table 1 as well. The average power spectrum corresponding to each group is shown in Figure 4. Notice that we will refer to the observations with the strongest BLN on the HB we studied above as group A. 

We investigated the relative frequency changes of the kHz QPOs due to the flux variation on the BLN time scales for group B, C, and D, similar to our study of group A. We repeated the analysis in the same way as what we performed to the observations of group A. In the frequency range 0.5--16 Hz, the average frequency drifts of $\nu_1$ and $\nu_2$ are $3.3\pm0.7$ Hz and $0.5\pm0.5$ Hz for group B, $2.5\pm0.5$ Hz and $-0.7\pm0.4$ Hz for group C, and $3.4\pm1.2$ Hz and $0.9\pm1.1$ Hz for group D, respectively.  With these average drifts, we also calculated the variations of frequency separation $\Delta\nu$ of the twin kHz QPOs. They are $2.8\pm0.9$ Hz, $3.2\pm0.6$ Hz, and $2.5\pm1.7$ Hz for B, C and D group, respectively. As an example, the results corresponding to group D is shown in Figure 5.

Observations show that the BLN increases in frequency when \mbox{Sco X$-$1} evolves from the HB to the NB, followed by the occurrence of the NBO, fading away before the occurrence of the NBO (see van der Klis et al. 1997). A summary of the results obtained for the BLN on the HB and the upper NB, as well as the results obtained for the NBO on the NB, are shown in Figure 6. It is interesting to see that, in response to an increase of the X-ray count rate on the BLN and the NBO time scales, the kHz QPO drifts are in opposite sign and the change of $\Delta\nu$ differ significantly. This indicates that the BLN and the NBO are of distinct origins.

\section{DISCUSSION}
We have analyzed {\it RXTE} observations of \mbox{Sco X$-$1} when the BLN and the twin kHz QPOs co-exist on the HB and the upper NB. We found that the frequencies of the twin kHz QPOs vary on the BLN time scales systematically. Associated with the BLN flux variation on the HB, the upper kHz QPO frequency varies by $1.8\pm0.4$ Hz, positively correlated with the BLN flux variation. On the other hand, from the HB to the very upper NB in sequence of decreasing BLN strength, the lower kHz QPO frequency is found to vary by $4.4\pm0.6$ Hz to $2.5\pm0.5$ Hz, correlating to the flux variation associated with the BLN in a positive way. Consequently, the twin kHz QPO peak separation decreases by $2.6\pm0.7$ Hz to $3.3\pm0.6$ Hz, opposite the flux variation due to the BLN. It is worth noting that on time scales longer than the BLN time scale, there is no significant correlation between the kHz QPO frequencies and the X-ray count rate among the observations. This is  exactly the same `parallel track' phenomenon seen in atoll sources (see the review of van der Klis 2006). The reason for this phenomenon might be that the accretion flow is composed of a disk flow and a halo flow (e.g., Kaaret et al. 1998) and only the disk flow determines the kHz QPO frequency (van der Klis 2001)

Both the upper and the lower kHz QPO frequencies are positively correlated with the flux variation on the BLN time scales on the HB. The correlation between the upper kHz QPO frequency and the BLN flux is opposite in sign to that between the upper kHz QPO frequency and the NBO flux. Besides, the scale of the frequency drift of the upper kHz QPO due to the flux variation on the time scale of the BLN on the HB and the upper NB, i.e., $1.8\pm0.4$ Hz, is much smaller than the $\sim$ 22 Hz drift in response to the NBO flux variation on the lower NB. This provides evidence that the BLN on the HB and the upper NB is probably of a different origin from the NBO. On the lower NB, the frequency of the upper kHz QPO is anti-correlated with the count rate variation on the NBO time scale (Yu et al. 2001). This is interpreted as that it probably arises from the effect of the radiative stresses on the inner disk edge which modulates the orbital frequency in the accretion flow at the NBO frequency. On the BLN time scales, the frequencies of the twin kHz QPOs are both positively correlated with the count rate, which suggest that the frequency drifts arise from the variation of the mass accretion rate. 

Associated with the flux variation on the time scale of the BLN, $\Delta\nu$ decreases with increasing $\nu_2$ by a few hertz. This behavior  is qualitatively consistent with the overall decreasing trend of $\Delta \nu$ with $\nu_2$ for neutron star LMXBs when the upper kHz QPO is about 910--980 Hz (see the plot in Stella \& Vietri 1999).  This hints that the same mechanism causes the kHz QPO frequency variations on the BLN time scales and on the longer time scales  that the neutron star LMXBs move along the tracks in the color-color diagram. This also supports that the BLN corresponds to the variation of the mass accretion rate in the accretion flow. Further support of this idea is from the comparison of the variability amplitude and the corresponding frequency shift on both the BLN time scales and the time scale a Z source evolves along the tracks in the CD. The {\it rms} amplitude of the BLN is on the scale of about 1\%. This is associated with a kHz QPO frequency drift of a few Hz. The relation between the flux variation and the kHz QPO frequency drift is quantitatively consistent with that of a Z source evolves along the tracks in the CD. The frequency range of the kHz QPOs of a few hundred Hz and the corresponding flux variation is on the scale of 50\% for a Z source to tracing the HB and the NB.

In neutron star LMXBs, the accretion geometry might consist of a disk flow and a radial flow (e.g., Kaaret et al. 1998; van der Klis 2001; Yu et al. 2004). The variation of the mass accretion rate in a disk flow or in a radial flow leads to different source behaviors, which helps determine the origin of the BLN. The frequencies of the HBO and the kHz QPOs on the HB are correlated with the X-ray flux in several Z sources (Lewin et al. 1992; Kuulkers et al. 1994; Kuulkers \& van der Klis 1996; van der Klis 1996a,b; Wijnands et al. 1996, 1997, 1998; Jonker et al. 1998; Dieters \& van der Klis 2000; Homan et al. 2002). One possible explanation is that the HBO frequency is probably determined by the mass accretion rate through the inner disk  (Wijnands et al. 1996). On the other hand, the NBO is different from the HBO. The upper kHz QPO frequency is anti-correlated with the NBO flux in \mbox{Sco X$-$1} (Yu et al. 2001), which suggests that the NBO probably originates from a radial flow, as in the model of Fortner et al. (1989) or Titarchuk (2001). On the BLN time scales we see a positive correlation between the kHz QPO frequency and the X-ray flux.  This suggests that the BLN is very likely associated with the variation of the mass accretion rate in the disk flow, similar to the HBO, rather than the variation of the mass accretion rate of the radial flow.

As shown in van der Klis et al. (1997), the NBO in \mbox{Sco X$-$1} seems to emerge from the BLN on the NB. However, the BLN and the NBO in \mbox{Sco X$-$1} may be of distinct origins (van der Klis 2007, private communication). Our study shows clearly that the BLN and the NBO are  different. Di Salvo et al. (2001) showed a similar phenomenon in an atoll source \mbox{4U 1728$-$34} in which a broad noise component disappeared immediately before a narrow peaked component, similarly to the NBO in the Z sources, emerged around 7 Hz.  The broad noise component and the narrow peaked component observed in the atoll sources may corresponds to the BLN and the NBO in \mbox{Sco X$-$1}, and of similar origins, respectively. 

It is worth noting that the frequency separation $\Delta\nu$ of the twin kHz QPOs varies on time scales of seconds or shorter (e.g., Yu et al. 2001; this letter). On the NBO time scale when the \mbox{Sco X$-$1} tracks on the NB, the lower kHz QPO seems to disappear and the upper kHz QPO decreases by at least 22 Hz when the NBO flux increases from low to high (see Yu et al. 2001); on the BLN time scales when the \mbox{Sco X$-$1} is on the upper NB and on the HB, the frequency separation varies by about a few Hz. These suggest that the frequency separation measured on longer time scales is probably smaller than the actual separation by at least 22 Hz on the NB, and no less than a few Hz on the upper NB and the HB. Therefore, with increasing upper kHz QPO frequency in the range from 920 Hz to 1090 Hz, the $\Delta\nu$ doesn't decrease as steeply as measured on longer time scales at least for \mbox{Sco X$-$1}. Since this trend has been used to constrain models (Stella \& Vietri 1999; Zhang et al. 2004), while these models predict much steeper trend towards higher frequencies than the observed, further observational studies of $\Delta\nu$ at highest kHz QPO frequencies are crucial. 

\section{Conclusion}
We have performed a study of f the band-limited noise (BLN) in \mbox{Sco X$-$1}. We conclude 
\begin{itemize}
\item Both the lower and the upper kHz QPO frequencies are positively correlated with the flux variation taking place on the BLN time scales, roughly quantitatively consistent with the kHz QPO frequency vs. the X-ray flux correlation hold along the Z track.
\item The frequency separation between the two kHz QPOs decreases with increasing kHz QPO frequencies on the BLN time scales by a few hertz, consistent with the overall trend held on longer time scales at the same kHz QPO frequency range. 
\item Based on the above properties of the BLN which is different from those of the NBO, we conclude that the BLN likely corresponds to the variation in the mass accretion rate of the disk accretion flow.  
\item The results obtained for the BLN and the NBO in \mbox{Sco X$-$1} suggest that further observational studies of $\Delta\nu$ at highest kHz QPO frequencies are crucial for testing kHz QPO models. 
\end{itemize}

\acknowledgments
We would like to thank the anonymous referee for kindness and very useful suggestions which makes this work complete. WY appreciate very useful discussions with Michiel van der Klis of the University of Amsterdam. This work was supported in part by the National Natural Science Foundation of China (10773023, 1083302, 10773004,10603002), the One Hundred Talents project of the Chinese Academy of Sciences, Shanghai Pujiang Program (08PJ14111), the National Basic Research Program of China (2009CB824800), and the starting funds of the Shanghai Astronomical Observatory. The study has made use of data obtained through the High Energy Astrophysics Science Archive Research Center Online Service, provided by the NASA/Goddard Space Flight Center.

\clearpage

\begin{table}
\begin{tabular} {ccccc}
Observation ID   & Group A & Group B & Group C & Group D\\\hline
10056-01-01-00&all&&&\\
10056-01-01-01&all&&&\\
10056-01-01-03&all&&&\\
10061-01-01-01&all&&&\\
20053-01-01-00&all&&&\\
10056-01-01-02 & &   1  & 2 & - \\ 
10056-01-02-00 & & 1, 2 & - & - \\ 
10056-01-02-01 & &1 & 2, 3 & - \\ 
10056-01-02-03 & & - & - & 1, 2, 3 \\ 
10056-01-02-04 & &  - & - & 1, 2 \\ 
10056-01-03-01 & &  - & 3 & 2 \\ 
10056-01-03-02 & &  3 & 1, 2 & -  \\ 
10061-01-02-00 & &- & 3 & 1, 4 \\ 
20053-01-01-01 & & 1 & 2, 3 & 4 \\ 
20053-01-01-04 & & 2 & 1, 3 & - \\ 
20426-01-02-00 & &- & 1 & -\\
20426-01-02-01 & &- & - & - \\
30053-01-01-00 & &2, 3 &1, 4, 5, 6, 7, 8 & -\\
30053-01-08-00 & &- & 2 & 1\\
40020-01-01-01 && - & 5 & - \\
40020-01-01-07 & &- & 1, 2, 3 & - \\
40020-01-03-00 & &1 & 2, 3 & -\\
40020-01-03-01 & &- & 1 & - \\
40706-02-08-00 & &- & - & 1 \\
40706-02-09-00 & &- & - & 1\\
40706-02-10-00 & &- & - & 1, 2, 3, 4 \\
40706-02-12-00 & &- & - & 1 \\
70014-01-01-00 & &- & - & 1 \\
91012-01-01-03 & &1 & - & - \\
\end{tabular}
\caption{{The list of segments and observations corresponding to group A, B, C and D.  The segment number tells the segment sequence in the observation. `-' means no segment of the observation was selected.}}
\end{table}

\clearpage

\begin{figure}
\plotone{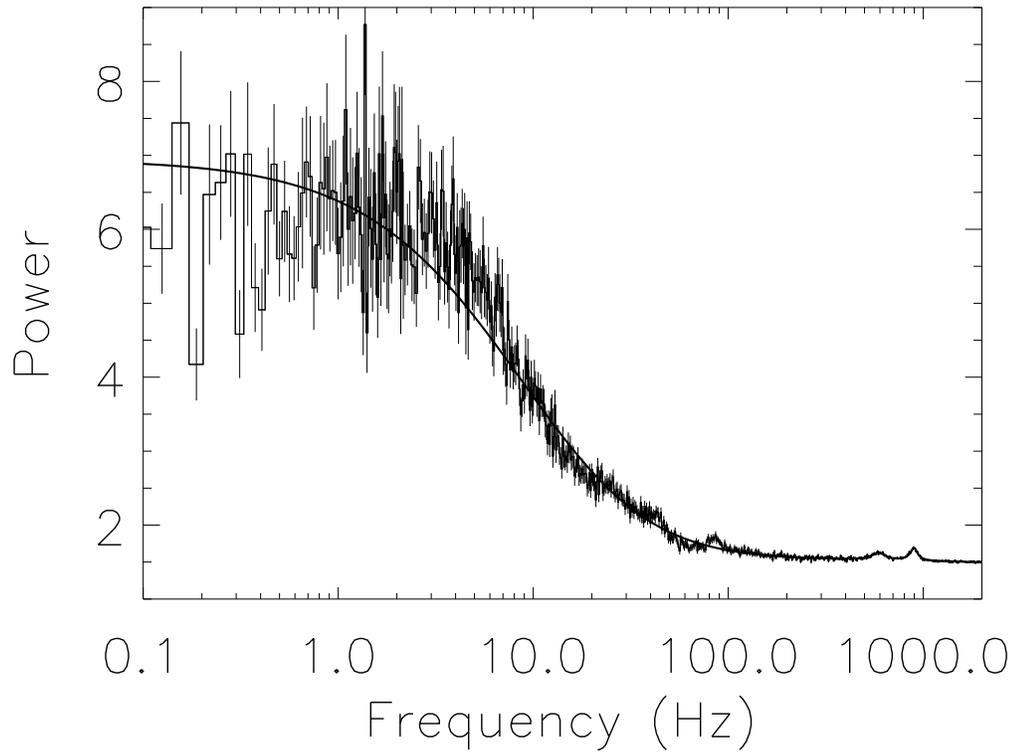}
\caption{The average power spectrum obtained from the Bin mode data during the observation 10056-01-01-01. It shows that the flat-topped BLN in the frequency range 0.5$-$4.0 Hz and the twin kHz QPOs are the major variability components of \mbox{Sco X$-$1} on the HB. An example of the power spectral fit with the model composed of a low frequency Lorentzian component, twin kHz Lorentzian components, and a constant plus a power-law noise component is shown as a solid line. }
\end{figure}

\begin{figure}
\epsscale{.8}
\plotone{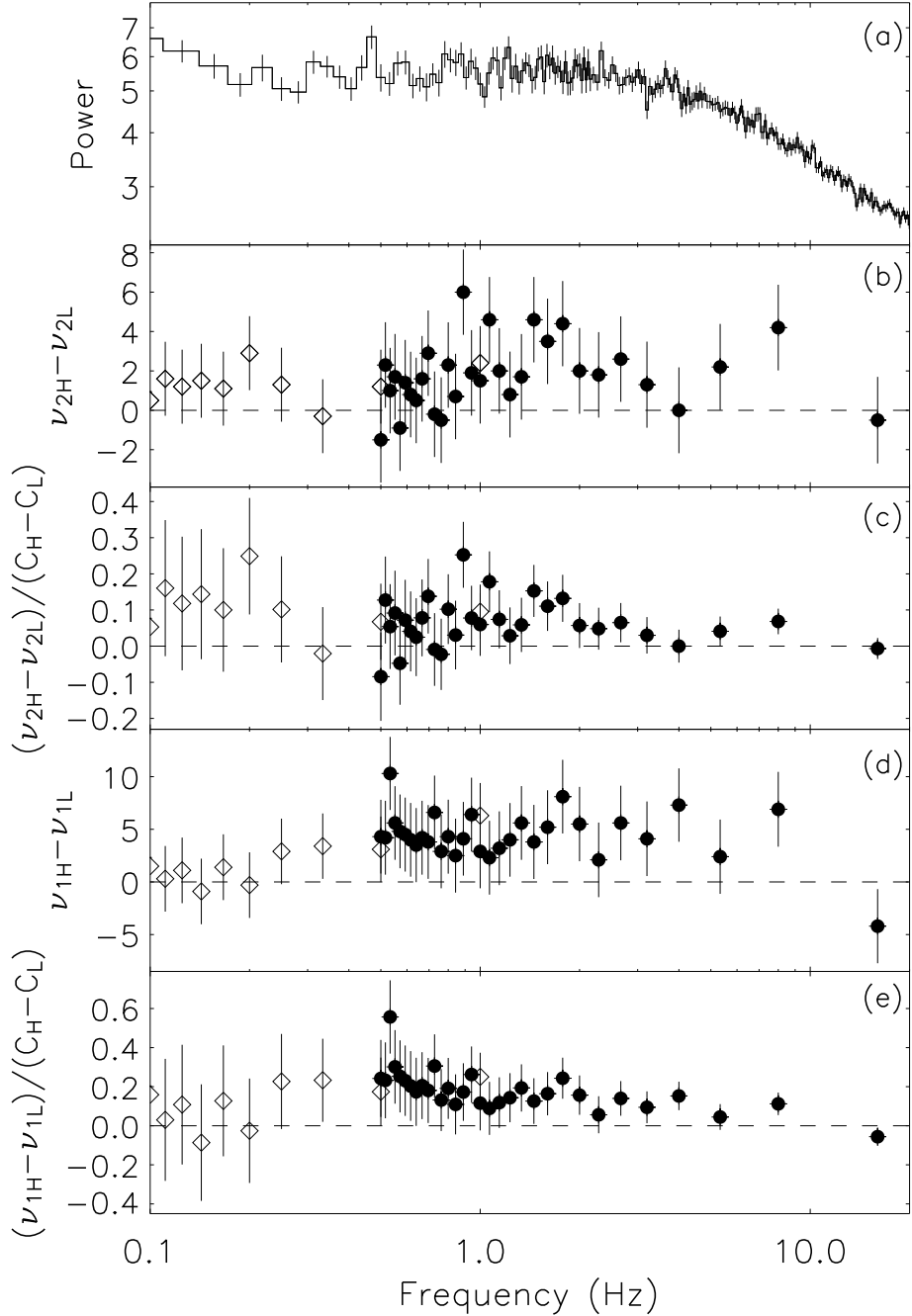}
\epsscale{1}
\caption{{Comparison between the high count rate samples and the low count rate samples with the observations of \mbox{Sco X$-$1} on the HB (group A). The filled circles and diamonds represent results from the analysis of $N*1/32$ s segments and  $M*0.5$ s segments, respectively. From top to bottom, (a) the average power spectrum showing strong BLN; (b) the frequency shifts of the upper kHz QPO; (c) the frequency shifts of the upper kHz QPO over the average counts in  1/32 s intervals; (d) the frequency shifts of the lower kHz QPO. ; (e) the frequency shifts of the lower kHz QPO over the average counts in 1/32 s intervals. Notice that the frequency shifts for the lower kHz QPO are larger than that of the upper kHz QPO by about 2 Hz on the BLN time scales.}}
\end{figure}

\begin{figure}
\plotone{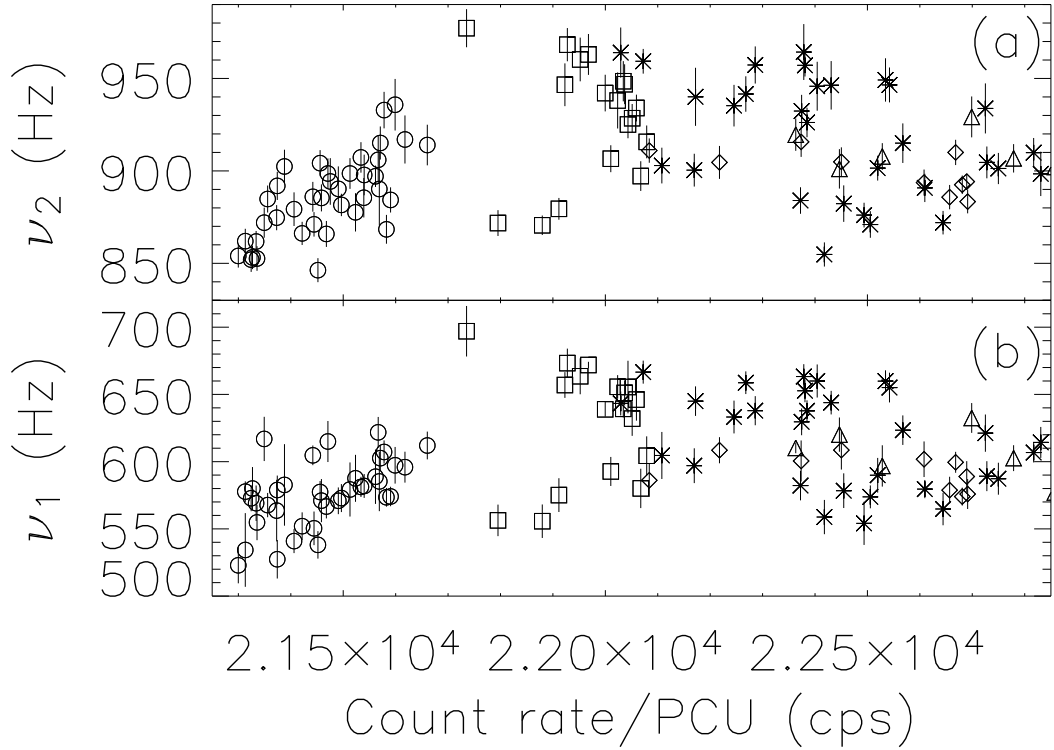}
\caption{ {The upper kHz QPO frequency vs. the X-ray count rate per CPU (upper panel) and the lower kHz QPO frequency vs. X-ray count rate per CPU (lower panel) on the time scale of 256 s in the five observations of \mbox{Sco X$-$1} on the HB (crosses: 10056-01-01-00; diamonds: 10056-01-01-01; triangles: 10056-01-01-03; squares: 10061-01-01-01; circles: 20053-01-01-00.). } }
 \end{figure}

\begin{figure}
\plotone{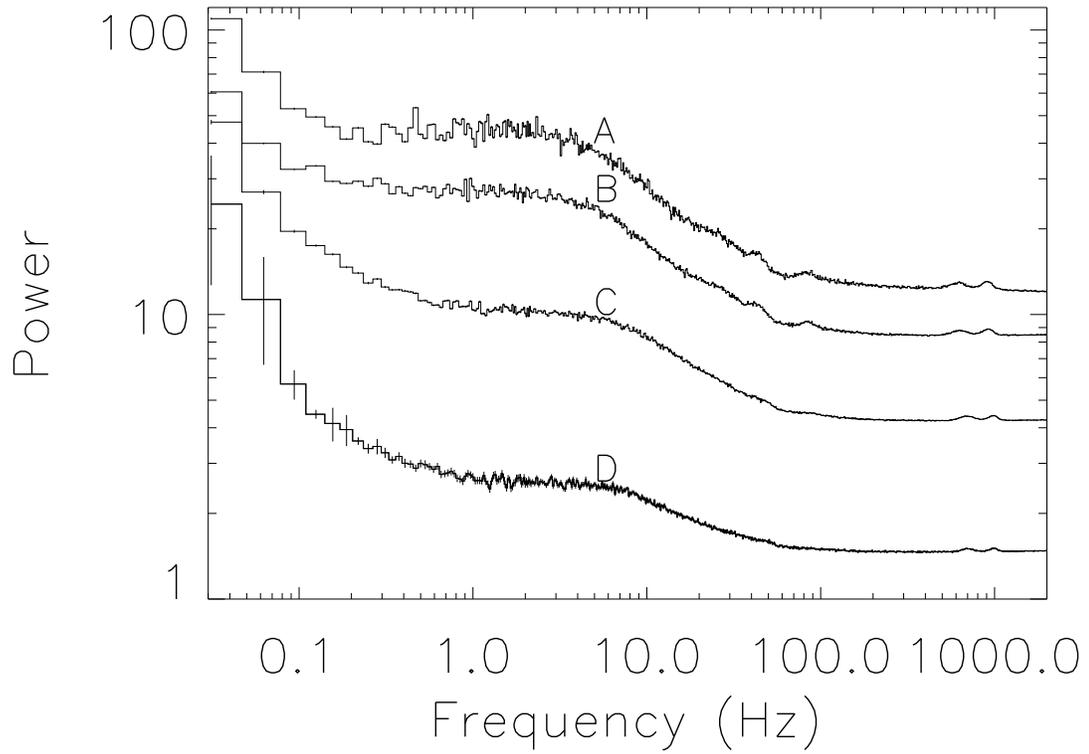}
\caption{{The average power spectra of group A, B, C, and D which trace the BLN component from the HB to the upper NB, corresponding to the observations with 0.5-4 Hz average power in the ranges A) larger than 5.0, B) 4.0--5.0, C) 3.0--4.0 and D) 2.0--3.0, respectively. To better show the power spectral evolution, the power spectra corresponding to `A', `B', and `C' are increased by 8, 6, and 3 times, respectively.} }
\end{figure}

\begin{figure}
\epsscale{.8}
\plotone{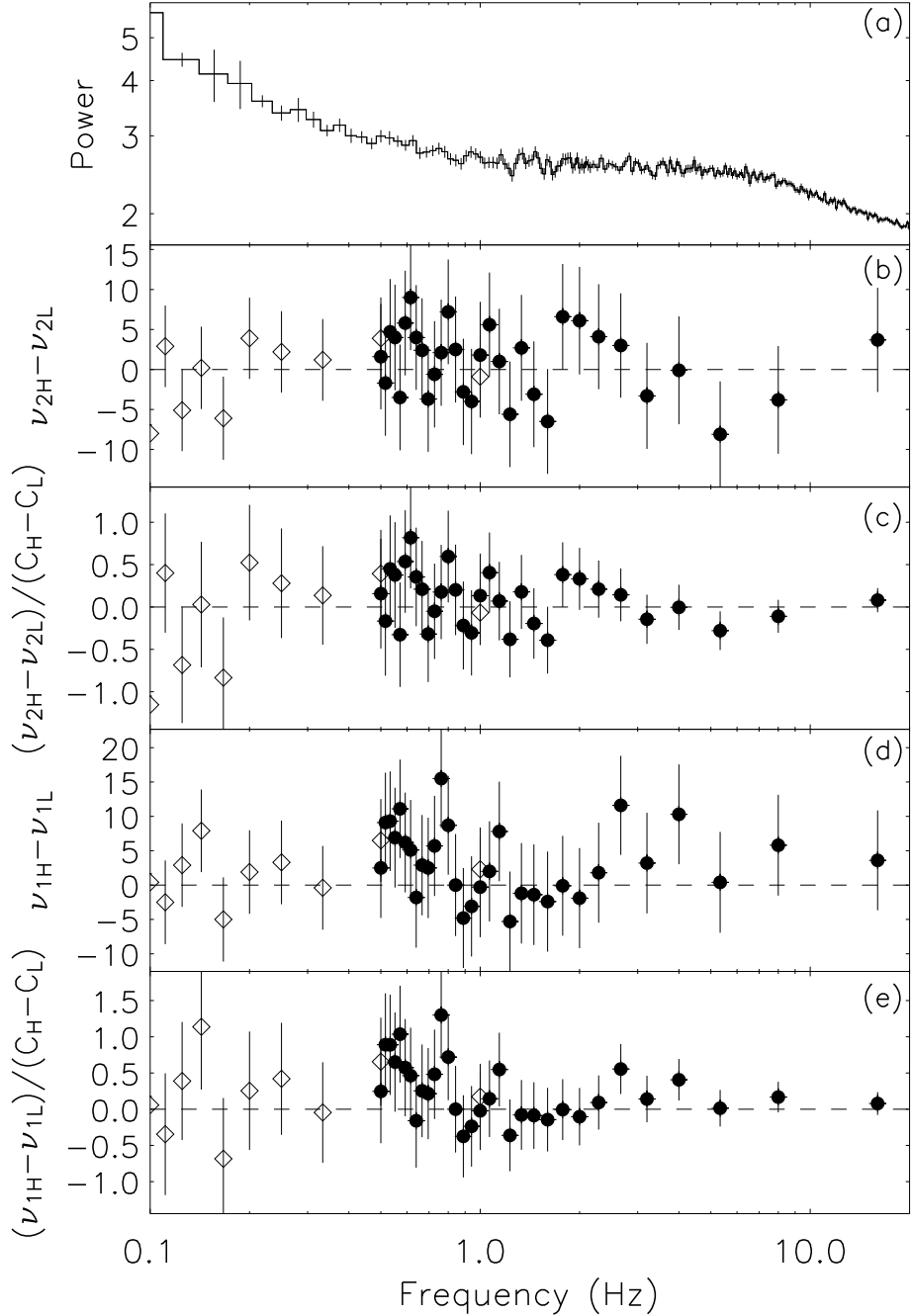}
\epsscale{1}
\caption{{Comparison between the high count rate samples and the low count rate samples for group D. The filled circles represent results from $N*1/32$ s segments and the diamonds represent results from $M*0.5$ s segments. From top to bottom, (a) the average power spectrum showing the BLN; (b) the frequency shifts of the upper kHz QPO; (c) the frequency shifts of the upper kHz QPO over the average counts in 1/32 s intervals; (d) the frequency shifts of the lower kHz QPO. (e) the frequency shifts of the lower kHz QPO over the average counts in 1/32 s intervals. } }
\end{figure}

\begin{figure}
\plotone{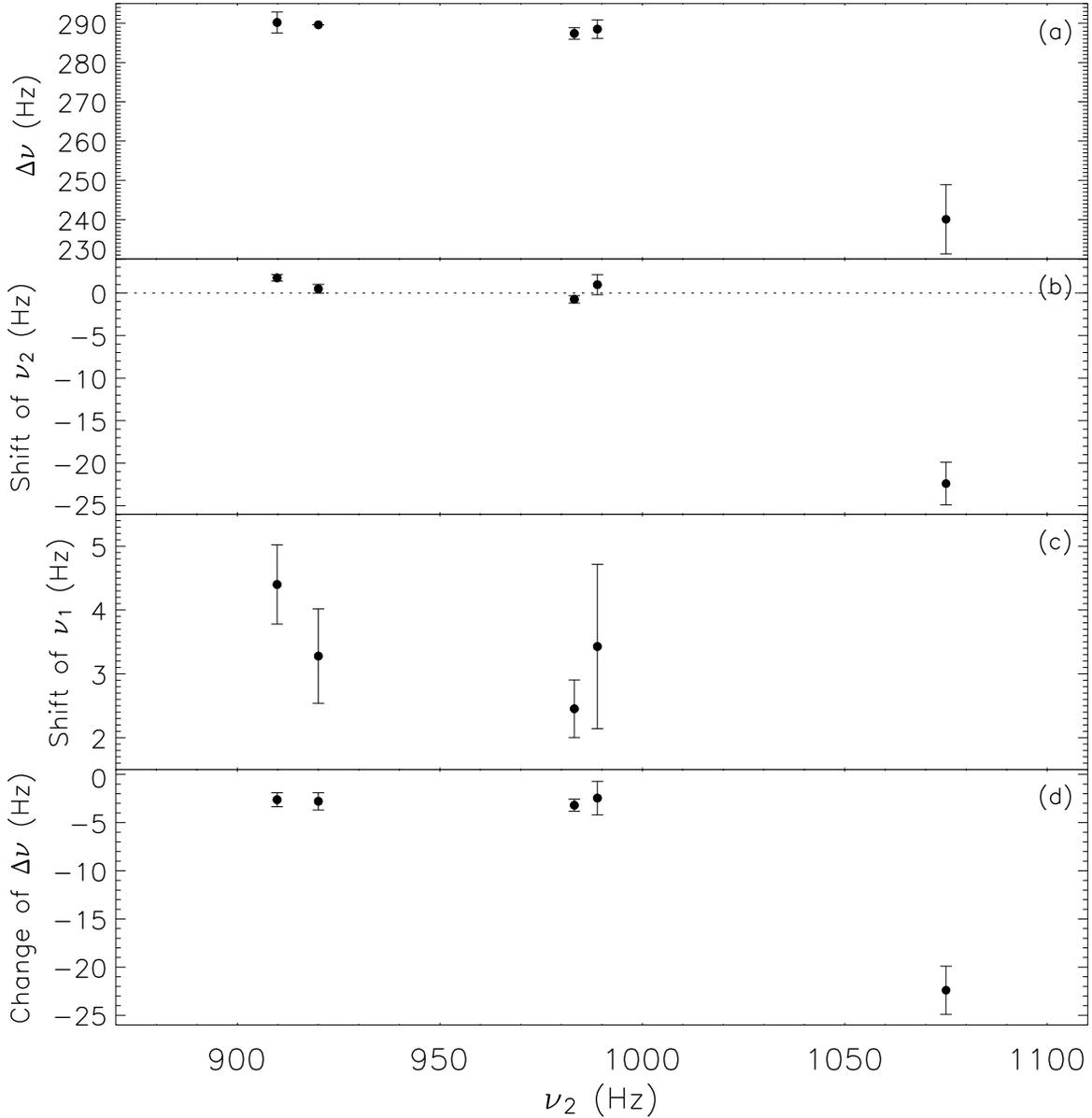}
\caption{{From the HB to the NB, the average frequency separation of the twin kHz QPOs (upper panel),  the frequency drifts of the kHz QPO frequencies (middle panel) and the variations of their frequency separations (lower panel) between the low count rate samples and the high count rate samples on the BLN or the NBO time scales. From left to right, the  data points correspond to group A, B, C, D, and those on the NB with strong NBO (Yu et al. 2001), respectively. From top to bottom, against the upper kHz QPO frequencies measured in the average power spectra: (a) the frequency separations of the twin kHz QPOs (the NB data is estimated from the observation 10056-01-04-02); (b) the frequency drifts of $\nu_1$  between the low count rate samples and the high count rate samples; (c) the frequency drifts of $\nu_2$; (d) the change of the frequency separation ( the NB data is from Yu et al. 2001, assuming that the lower kHz QPO frequency is fixed during an NBO cycle). }}
\end{figure}

\end{document}